\documentclass[aps,prd,superscriptaddress,floats,tightenlines,balancelastpage,floatfix]{revtex4}
\input epsf
\usepackage{amsmath,amssymb}
\usepackage{epsfig}
\usepackage{graphicx}

\usepackage{epstopdf}

\begin{document}

\title{Spectral Distortion in a Radially Inhomogeneous Cosmology}

\author{R. R. Caldwell}
\author{N. A. Maksimova}
\affiliation{Department of Physics \& Astronomy, Dartmouth College, 6127 Wilder Laboratory, Hanover, NH 03755 USA}

\date{\today}

\begin{abstract}

The spectral distortion of the cosmic microwave background blackbody spectrum in a radially inhomogeneous spacetime, designed to exactly reproduce a $\Lambda$CDM expansion history along the past light cone, is shown to exceed the upper bound established by COBE-FIRAS by a factor of approximately $3700$. This simple observational test helps uncover a slew of pathological features that lie hidden inside the past light cone, including a radially contracting phase at decoupling and, if followed to its logical extreme, a naked singularity at the radially inhomogeneous Big Bang.

\end{abstract}

\maketitle

\section{Introduction}

Is the Universe playing fair with us? Are the laws of physics and the structure of space-time the same everywhere? It is a fundamental tenet of the Standard Cosmological Model that the answer is yes. Yet the difficulty of explaining the physics of cosmic acceleration forces a new scrutiny of many of our most cherished assumptions. If the structure of space-time is not the same everywhere, if in fact we occupy a privileged location in space and time at the center of a spherical bulge of matter and curvature, then it may be possible to explain a vast catalog of observational data without the need to invoke new physical effects such as dark energy \cite{Celerier:1999hp,Tomita:2000jj,Zibin:2008vk}.

In this article we consider the consequences if the answer to the above questions is no. In particular, we consider a toy model of the Universe containing no dark energy and invoking no new gravitational physics. Instead, the space-time is filled by spherically-symmetric, comoving shells of pressureless dust, according to the Lemaitre-Tolman-Bondi (LTB) metric \cite{Lemaitre:1933gd,Tolman:1934za,Bondi:1947av}. This particular model is characterized by an inhomogeneous Big Bang surface. The profile of the mass density is not uniform, consisting of a slight bulge near the origin, in contrast to LTB models that carve out a significant, Gpc radius void, e.g. Refs.~\cite{Garfinkle:2006sb,Alexander:2007xx,February:2009pv}. Here, the density varies by just a few percent between the origin and the physical Hubble distance, in such a way that an observer located at the origin will infer an expansion history that matches the Standard Cosmological Model even though there is no cosmological constant and no cosmic acceleration \cite{Celerier:2009sv,Kolb:2009hn,Dunsby:2010ts}.

\begin{figure}[h]
\includegraphics[width=1\linewidth]{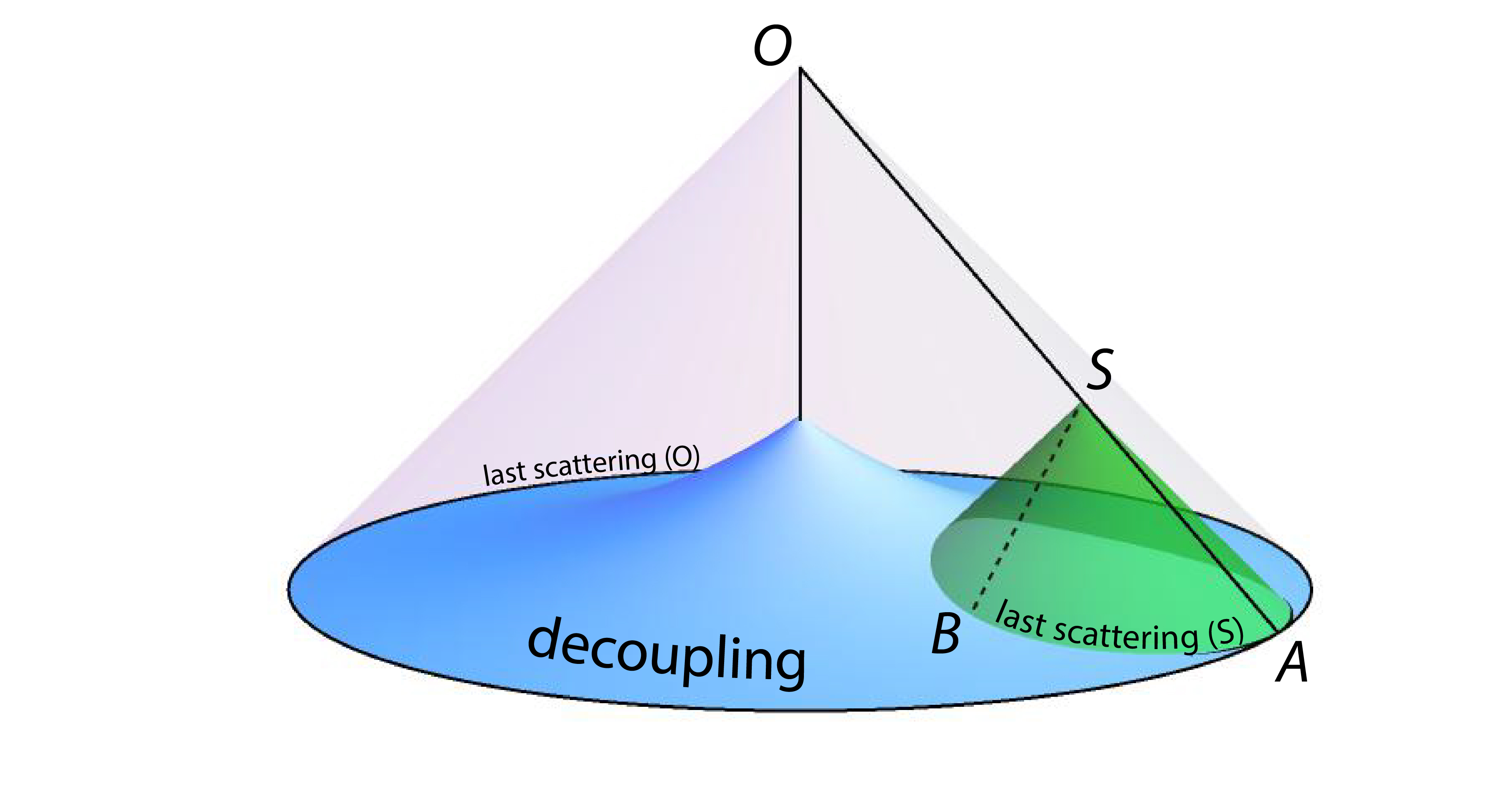}
\caption{The causal structure of our model space-time. Here and now is point $O$. The cuspy surface is the constant-energy-density decoupling surface. The intersection of the past light cone of $O$ with the decoupling surface yields the last scattering surface. One possible line of sight to last scattering is shown as the line $OA$. However, photons that originate within our past light cone may scatter off free electrons after reionization at point $S$ and merge with photons in our line of sight. Because the temperature of photons that originate on the last scattering surface of $S$ at $B$, for example, is in general different from those originating at $A$, the mixture produces a distortion of the line-of-sight blackbody.}
\label{fig:spacetime}
\end{figure}

Scattered light originating from within our past light cone, however, tells a different story. Cosmic microwave background (CMB) photons emitted at the epoch of decoupling within our past light cone encounter strong gravitational fields. Even though only a small fraction of the photons scatter off free electrons and into our line of sight, the resulting spectrum is no longer a blackbody. To help explain this process, a sketch of the light cone structure of our model space-time is presented in Fig.~\ref{fig:spacetime}. The temperature of secondary photons that originate deep inside our past light cone is in general different from those primary photons that start on the past light cone. The mixture of primary and secondary photons produces a distortion of the primary, line-of-sight blackbody. In this article we calculate the u-distortion of the CMB \cite{Goodman:1995dt,Stebbins:2007ve,Caldwell:2007yu}, which measures the degree of departure from a black body in terms of the width of the temperature distribution. Our main result, presented at the end of Sec.~\ref{sec:udist}, is that the u-distortion is approximately $3700$ times larger than the bound $u < 3.0 \times 10^{-5}$ (95\% CL) on spectral distortions set by COBE-FIRAS \cite{Mather:1993ij,Fixsen:1996nj}. This model is ruled out.

The strong u-distortion reveals a host of pathological features deep inside the past light cone. Geodesics that start on the radially inhomogeneous Big Bang surface from within our past light cone are not redshifted as expected. Instead, they are infinitely blueshifted, revealing a naked singularity. Even if we include only the portion of the space-time that extends back to the epoch of decoupling, we discover that a portion of the space-time is contracting. This means secondary photons that pass through the contracting region are blueshifted and gain energy, leading to the excessive u-distortion.

Cosmological scenarios built upon a radially inhomogeneous space-time have been explored extensively as an alternative to dark energy, in large part to explain the expansion history sampled by type 1a supernova luminosity distances and other classical tests of cosmology \cite{Romano:2007zz,Romano:2009qx,Romano:2009mr}. These models also predict excess velocities of large scale structure relative to CMB photons, which conflicts with observations of the CMB spectrum in the direction of hot, gaseous clusters, also known as the kinetic Sunyaev-Zeldovich effect \cite{Sunyaev:1980nv}. Hence, many of these cosmological scenarios are already tightly constrained if not ruled out \cite{GarciaBellido:2008nz,GarciaBellido:2008gd,Hunt:2008wp,Yoo:2010qy, Biswas:2010xm,Moss:2010jx,Yoo:2010ad,Zhang:2010fa,Caldwell:2010,Moss:2011ze,Bull:2011wi}. In Ref.~\cite{Zibin:2011ma}, the spectral distortion of the CMB is calculated for a general class of models with an inhomogeneous Big Bang. Although this class of models does not include the scenario presented in our work, the results are comparable.

The u-distortion, despite its similarity to the SZ effect, presents an opportunity to place tighter constraints on these models. With an eye towards future tests, our work shows that the u-distortion, or any such probe that samples the physics deep within the past light cone, can be used to test radial homogeneity.  
 
The outline of our article is as follows. In Sec.~\ref{sec:model} we present the toy model and describe our method of solving for geodesics. In Sec.~\ref{sec:udist} we evaluate the u-distortion. In Sec.~\ref{sec:pathology} we examine the pathologies of the space-time revealed by the u-distortion. We summarize our results in Sec.~\ref{sec:summary}.

\section{The Model}
\label{sec:model}

The framework of our cosmological model is a spherically symmetric space-time filled by comoving shells of pressureless dust. The radial profile of the mass density is not uniform, but varies with radius in such a way that the expansion history inferred from the luminosity distance-redshift relationship matches the predictions of the radially homogeneous $\Lambda$CDM model. In violation of the Cosmological Principle, we locate ourselves at a special point: the center of this space-time. 

The space-time that we describe is the Lemaitre-Tolman-Bondi (LTB) space-time \cite{Lemaitre:1933gd,Tolman:1934za,Bondi:1947av}, with line element
\begin{equation}
ds^2 = -dt^2 + \frac{R'^2(r,t)}{1+\beta(r)} dr^2 + R^2(r,t) d\Omega^2.
\end{equation}
A prime indicates the partial derivative with respect to $r$, ${}' = \partial/\partial r$, and an overdot is reserved for the partial derivative with respect to time, $\dot{} = \partial/\partial t$. The radial coordinate $r$ and scale factor $R$ have units of length. The quantity $\beta/r^2$ plays a role analogous to the curvature of comoving spatial sections in a Robertson-Walker space-time.

We extend a well-established recipe by Refs.~\cite{Mustapha:1998jb,Celerier:1999hp} but use the same model and notation as Kolb \& Lamb \cite{Kolb:2009hn} (hereafter KL) to build a space-time that is identical to the $\Lambda$CDM model on the past light cone in terms of the redshift dependence of the luminosity distance and the matter density. (A different recipe was presented in Ref.~\cite{Chung:2006xh}.)

\subsection{Gravitational Field Equations}

We start our analysis of the space-time from the Einstein field equations, $G^{\mu\nu} = \kappa T^{\mu\nu}$ with $\kappa = 8 \pi G_N$, which yield
\begin{eqnarray}
&& \frac{\dot R^2}{R^2} + 2 \frac{\dot R'}{R'}\frac{\dot R}{R} - \frac{\beta}{R^2} - \frac{\beta'}{R R'} =  \kappa \rho,
\label{eqn:gtt}\\
&& \frac{\dot R^2}{R^2} + 2 \frac{\ddot R}{R} -  \frac{\beta}{R^2}= 0.
\end{eqnarray}
The spherically-symmetric matter distribution is characterized by the stress-energy tensor $T^{\mu\nu} = \rho(r,t) u^\mu u^\nu$, where $u^\mu$ is the four-velocity of the comoving matter, i.e. $u^\mu = (1,0,0,0)$. The conservation of stress-energy $\nabla_\mu T^{\mu\nu}=0$ means $\partial_t  (\rho R' R^2)=0$, or $\rho \propto (R' R^2)^{-1}$, where the constant of proportionality describing the mass profile is independent of $t$. With foresight, we can choose the constant itself to be a derivative with respect to $r$, and call it $\alpha'$. Consequently, we can express the energy density in terms of the dimensionless quantity $\alpha'$ and metric coefficients $R,\, R'$
\begin{equation}
\kappa \rho(r,t) = \frac{\alpha'(r)}{R'(r,t) R^2(r,t)}.
\label{eqn:rho}
\end{equation}
Evolution equations for these metric coefficients are obtained by combining Eqs.~(\ref{eqn:gtt}-\ref{eqn:rho}), whereby
\begin{eqnarray}
\dot R &=& \sqrt{\beta + \alpha/R}, 
\label{eqn:EEeqn1}\\
\dot R' &=& \frac{\beta' + \alpha'/R - \alpha R' /R^2}{2 \dot R}.
\label{eqn:EEeqn2}
\end{eqnarray}
These may be used to determine the expansion rate in directions that are parallel and transverse to a line of sight from the center, $H_\parallel = \dot R'/ R'$ and $H_\perp = \dot R/R$ respectively. A parametric solution to these equations is given by
\begin{eqnarray}
R(r,t) &=& \frac{\alpha(r)}{2 \beta(r)}\left[ \cosh \eta(r,t)-1\right], 
\label{eqn:param1}\\
t-t_{BB}(r) &=& \frac{\alpha(r)}{2 \beta^{3/2}(r)} \left[ \sinh\eta(r,t) - \eta(t,r)\right],
\label{eqn:param2}
\end{eqnarray} 
where $t_{BB}(r)$ is the time of the Big Bang at a radial position $r$. The radial functions $\alpha$, $\beta$, and $t_{BB}$ remain to be determined in Appendix~\ref{appendixB}.

\subsection{Past Lightcone of Here and Now}

To build the same redshift dependence of the luminosity distance and the matter density as $\Lambda$CDM, we consider geodesics on the past light cone. For our space-time metric, radially-directed light rays satisfy the equation
\begin{equation}
\frac{dr}{dt} = - \frac{\sqrt{1 + \beta(r)}}{R'(r,t)}.
\label{eqn:drdt}
\end{equation}
The time rate-of-change of redshift, derived in Appendix~\ref{appendixA}, is
\begin{equation}
\frac{dz}{dt} = -(1+z)\frac{\dot R'(r,t)}{R'(r,t)}.
\label{eqn:dzdt}
\end{equation}
We define the solution to Eq.~(\ref{eqn:drdt}) as $t=\hat t(r)$, and hereafter use the hat to indicate quantities on the past light cone. Following Refs.~\cite{Mustapha:1998jb,Kolb:2009hn}, we exploit a coordinate freedom and rescale $r$ so that on the past light cone, $\widehat R' = R'(r,\hat t)=\sqrt{1+\beta(r)}$. In this case, $\hat t(r) = t_0 - r$ where $t_0$ is the present time at the origin. 

We follow the recipe outlined in Refs.~\cite{Mustapha:1998jb,Celerier:1999hp,Kolb:2009hn} for the physical prescription to determine the radial dependence of $\alpha,\, \beta$ and $\widehat R$. First, the luminosity distance in our spacetime is
\begin{equation}
d_L(z) = (1+z)^2 \widehat R(r,\hat t),
\end{equation}
so that in order to match to a $\Lambda$CDM scenario, for which
\begin{eqnarray}
d_L(z) &=& (1+z) \int_0^z \frac{dz'}{H_{\Lambda CDM}(z')} \\
H_{\Lambda CDM}(z) &=& H_0 \sqrt{\Omega(1+z)^3 + 1 - \Omega},
\end{eqnarray} 
we require that
\begin{equation}
\widehat R(z) =\frac{1}{1+z}\int_0^z \frac{dz'}{H_{\Lambda CDM}(z')}.
\label{eqn:Rofz}
\end{equation}
Second, we require that the energy density in matter evolves with redshift just as in the $\Lambda$CDM scenario, whereby 
\begin{equation}
\kappa \hat\rho(z) = 3 H_0^2 \Omega (1+z)^3.
\label{eqn:rhoz}
\end{equation}
Third, to maintain the same number density of sources per redshift and per solid angle as in the $\Lambda$CDM scenario, $dN/dz\, d\Omega$, then
\begin{equation}
\frac{dz}{dr} = (1+z) H_{\Lambda CDM}(z)
\label{eqn:dzdr}
\end{equation}
for the radial rate of change of redshift along the past light cone. 

In this study, we fix $\Omega = 0.3$ and $H_0 = 100 h$~km/s/Mpc with $h=0.7$. These values ensure an adequate fit to observational data along the past light cone, including type 1a supernova luminosity distances and the angular diameter distance of CMB anisotropies.

\begin{figure}[t]
\includegraphics[width=0.45\linewidth]{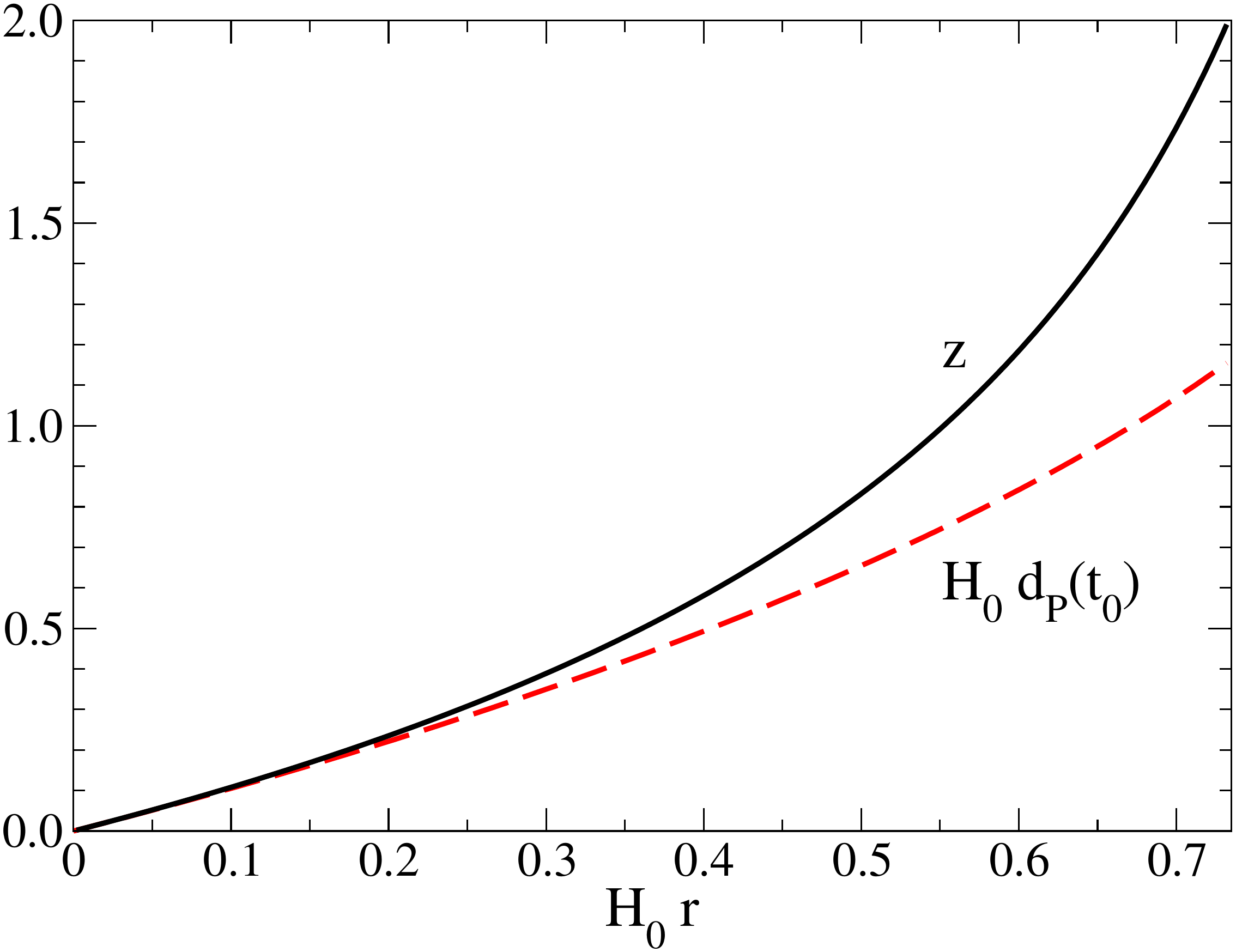}
\hspace{0.2cm}
\includegraphics[width=0.45\linewidth]{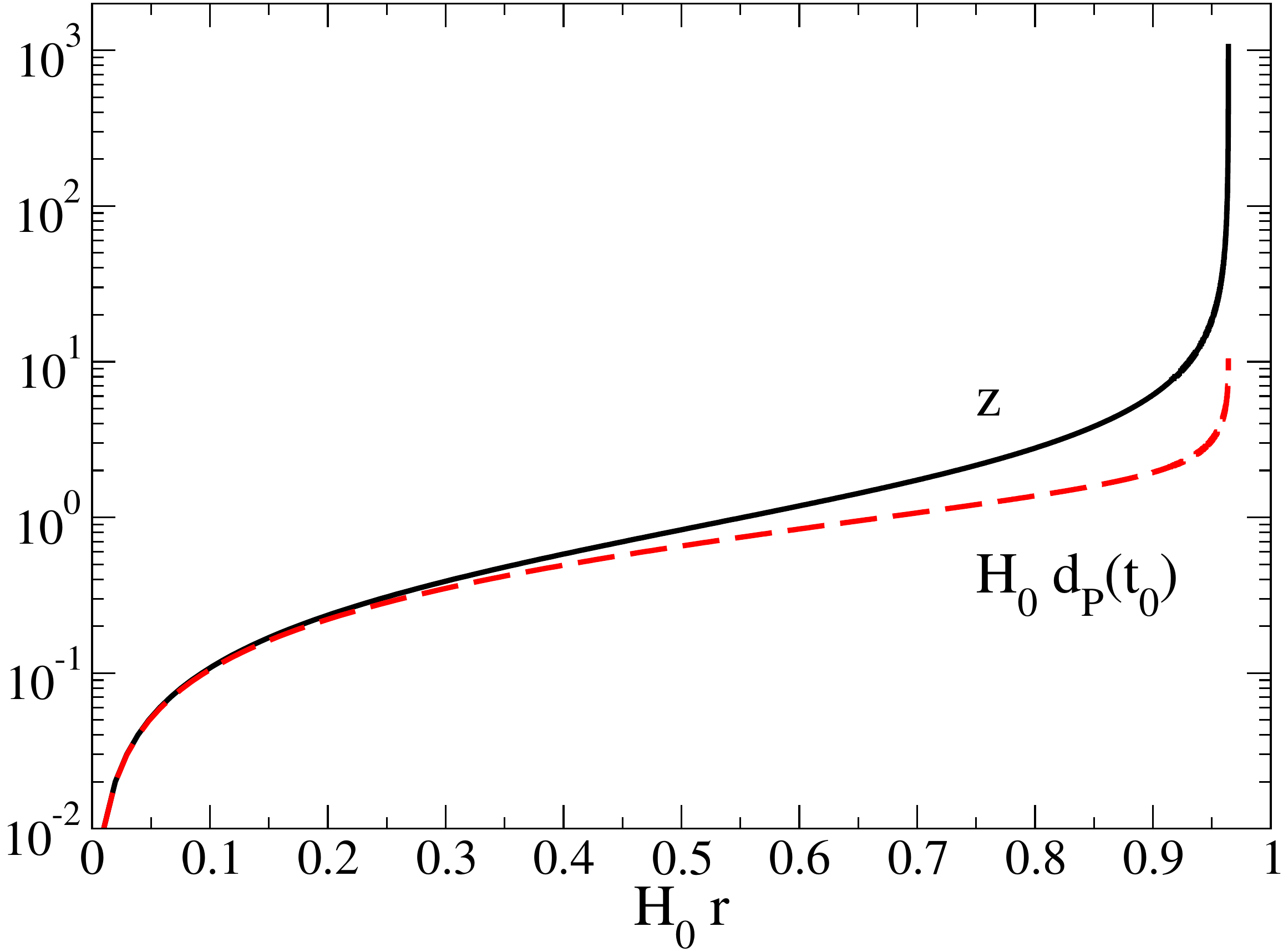}
\caption{The redshift and present-day proper distance are shown as a function of coordinate distance out to a redshift $z=2$ (left) and $z=1100$ (right) on the past light cone. The panel on the left is identical to Fig.~2 of KL \cite{Kolb:2009hn}.}
\label{fig:figKL2}
\end{figure}

\begin{figure}[h]
\includegraphics[width=0.45\linewidth]{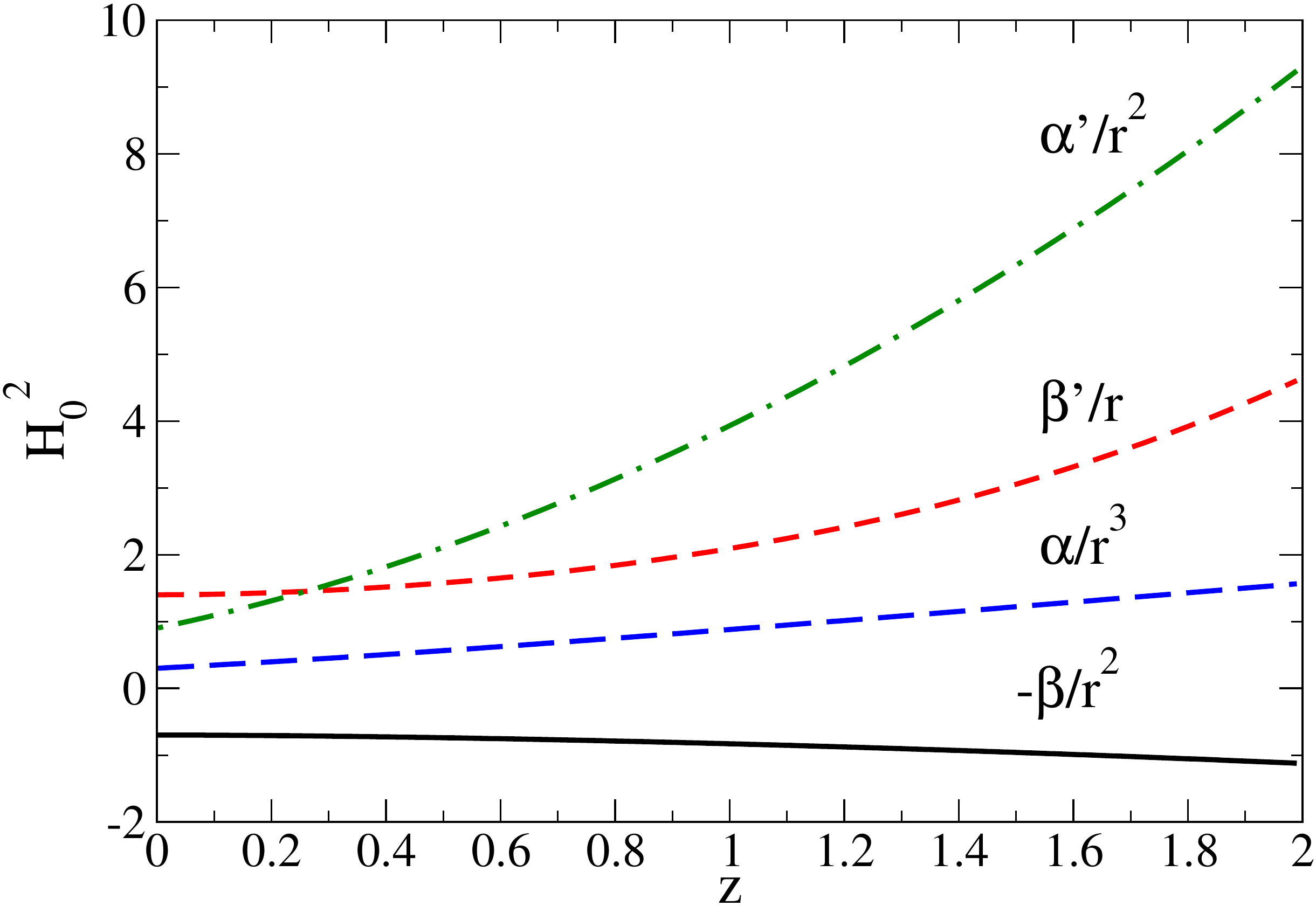}
\hspace{0.2cm}
\includegraphics[width=0.45\linewidth]{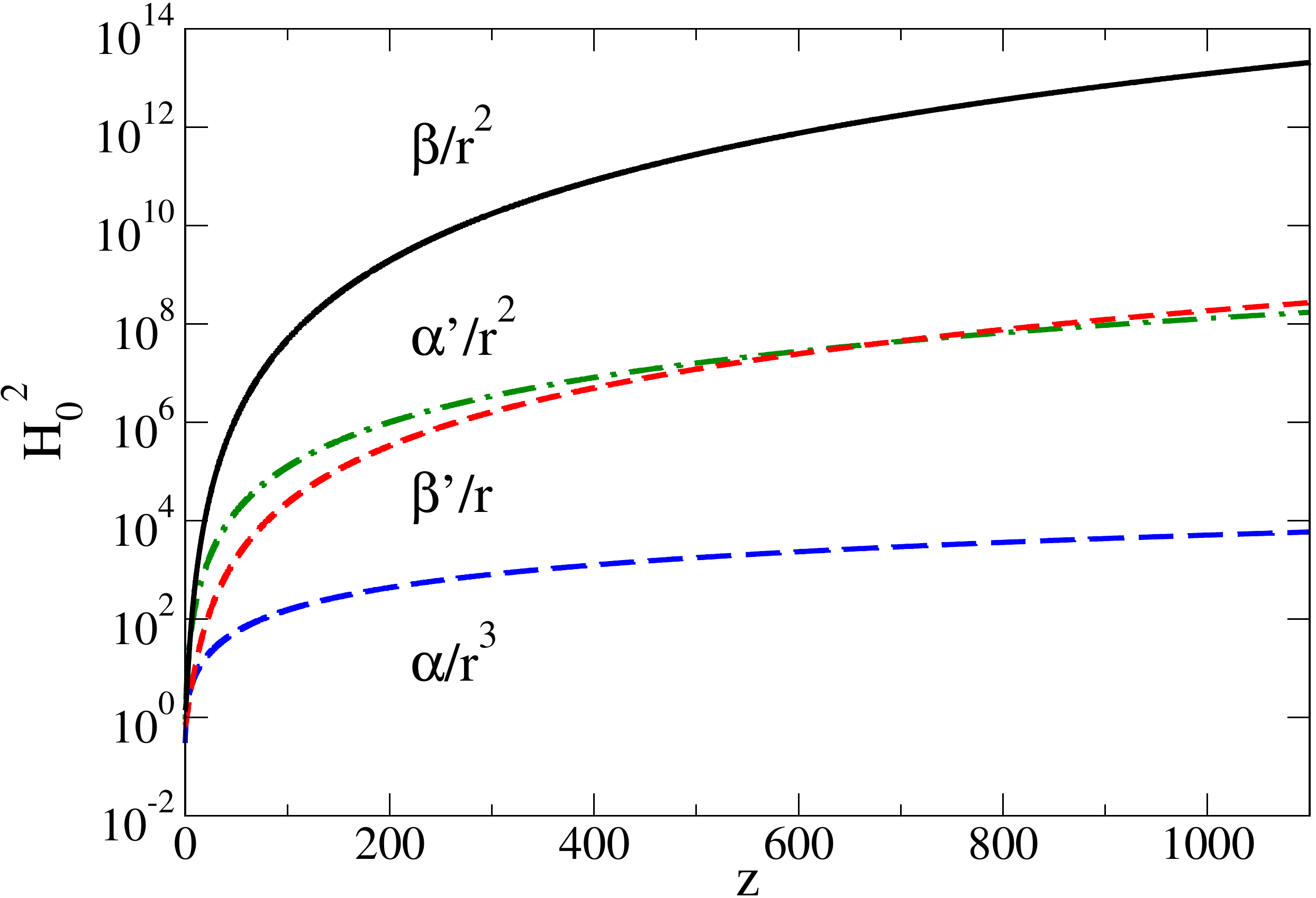}
\caption{The metric functions $\alpha,\,\beta$ and derivatives are shown as functions of redshift out to $z=2$ (left) and $z=1100$ (right). All quantities are shown in units of $H_0^2$. The panel on the left is identical to Fig.~3 of KL \cite{Kolb:2009hn}.}
\label{fig:figKL3}
\end{figure}

Equations for $\alpha,\,\beta$ and their derivatives as functions of the radial coordinate are derived in Appendix~\ref{appendixB}. We numerically integrate Eqns.~(\ref{eqn:Rofz}), (\ref{eqn:dzdr}), (\ref{Beqn:dVdz}-\ref{Beqn:lasteqn}). Results are exhibited in Figs.~\ref{fig:figKL2}-\ref{fig:figKL3}. The first panels in each show quantities out to a redshift $z=2$, which is sufficient to match the luminosity distance - redshift relationship charted out by type 1a supernovae. The second set of panels extends to $z=1100$ as will be needed to study the CMB. Hence, we have a procedure in place to construct all quantities on the past light cone, illustrated as the segment ${OA}$ in Fig.~\ref{fig:spacetime}.

\subsection{Inside the Past Light Cone }

\begin{figure}[h]
\includegraphics[width=0.35\linewidth]{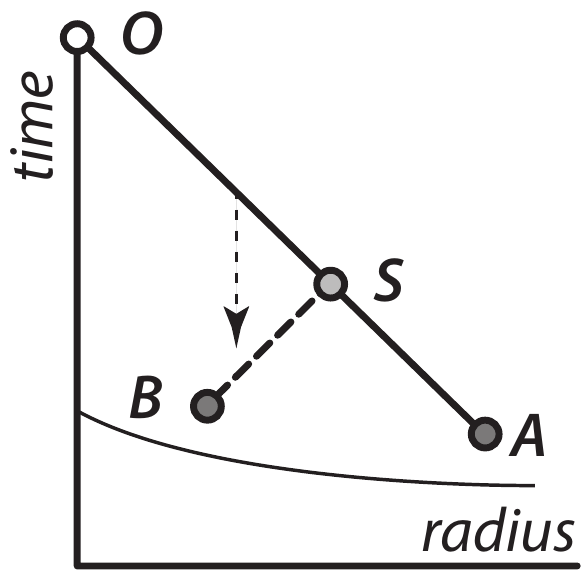}
\caption{The integration path is shown for solving the differential equations  for the metric variables $R,\,R'$ along geodesics inside the past light cone. The procedure outlined in Appendix~\ref{appendixB} is followed to obtain $\alpha,\,\beta$ and $R,\,R'$ along the past light cone, ${OA}$. Following a geodesic path from $S$ to $B$ requires $R,\, R'$ at each step. At a given value of $r$, we can evolve Eqs.~(\ref{eqn:EEeqn1}-\ref{eqn:EEeqn2}) down from the past light cone to the desired value of $t$, as illustrated by the dashed line with the arrowhead. (This is a similar recipe as illustrated in Fig.~4 of Ref.~\cite{Dunsby:2010ts}.)}
\label{fig:symbolic}
\end{figure}

Geodesics inside the past light cone are calculated using the equations described in Appendix \ref{appendixC}. In particular, we are interested in light rays that join the past light cone, as illustrated by the path ${BSO}$ in Fig.~\ref{fig:spacetime}. In practice, we follow these light rays backwards in time, starting at the present, here and now, moving back along the past light cone ${OS}$ until a particular scattering redshift is reached, and then fanning outwards along ${SB}$ at some angle relative to the path ${OS}$. The angle determines the angular momentum parameter $\ell$ defined in Eq.~(\ref{Ceqn:ell}). Then the set of differential equations (\ref{Ceqn:dzdt}-\ref{Ceqn:dudt}) are integrated, using Eq.~(\ref{Ceqn:ucon}) as a constraint, until we reach our destination at point ${B}$. Along this path, we also require $R,\,\dot R,\, R',\, \dot R'$ off the light cone. To evaluate these, we integrate Eqs.~(\ref{eqn:EEeqn1}-\ref{eqn:EEeqn2}) along a path of constant $r$, starting from the light cone at time $\hat t(r)$ down to the desired value of $t$. (See Fig.~\ref{fig:symbolic}.)

\begin{figure}[h]
\includegraphics[width=0.65\linewidth]{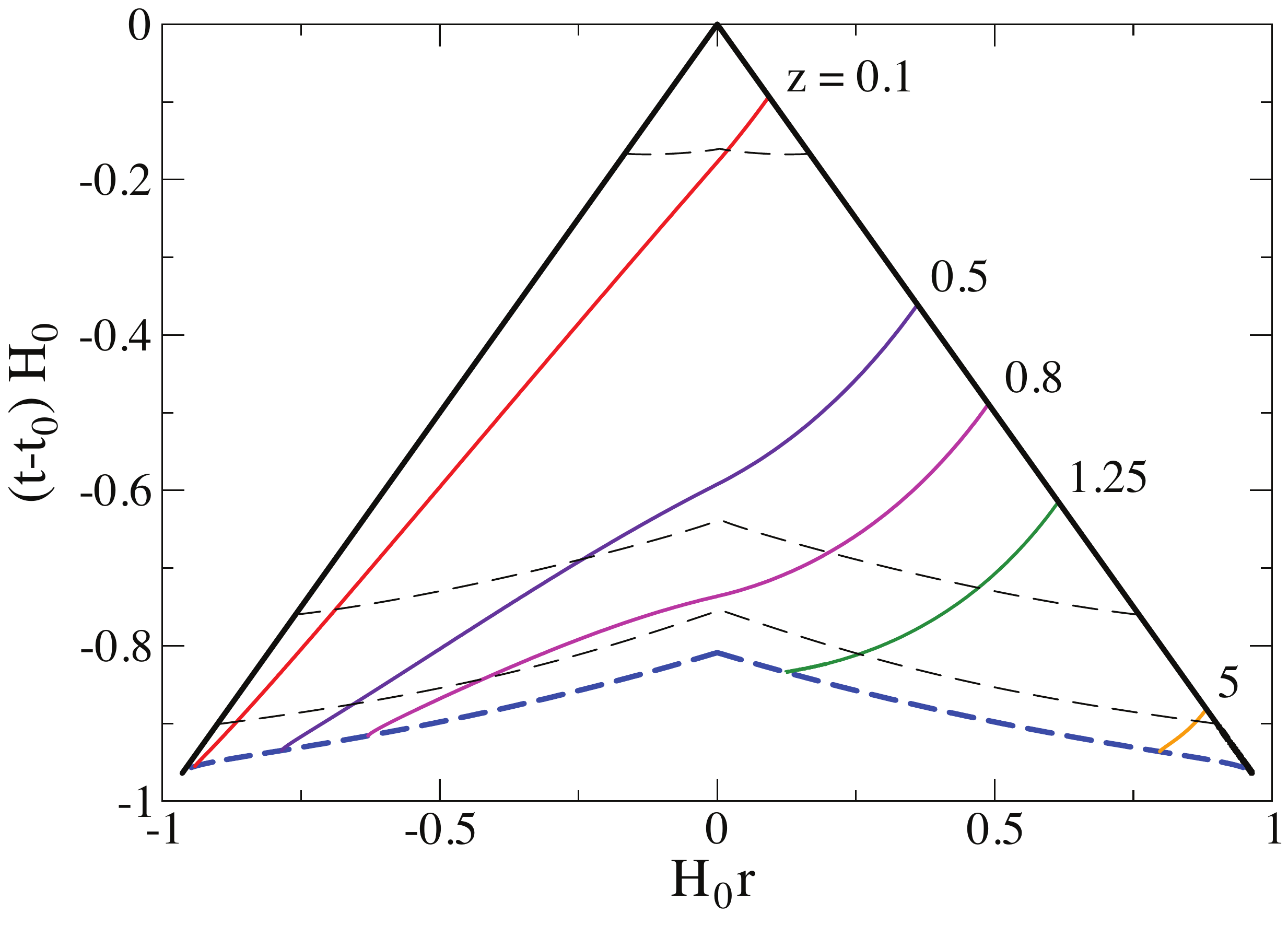}
\caption{Light rays that contribute to the spectral distortion of the CMB are shown on this space-time diagram. Radially-directed light rays that merge with the past light cone at a redshift $z$ are shown as solid lines. The past light cone is indicated by thick, dashed lines. The bottom, thick dashed line shows the inhomogeneous Big Bang surface. Thin dashed lines show surfaces of constant density, with values $1.5,\, 3,\, 300$ times the present-day density at the origin.}
\label{fig:180rescatter}
\end{figure}

\begin{figure}[h]
\includegraphics[width=0.5\linewidth]{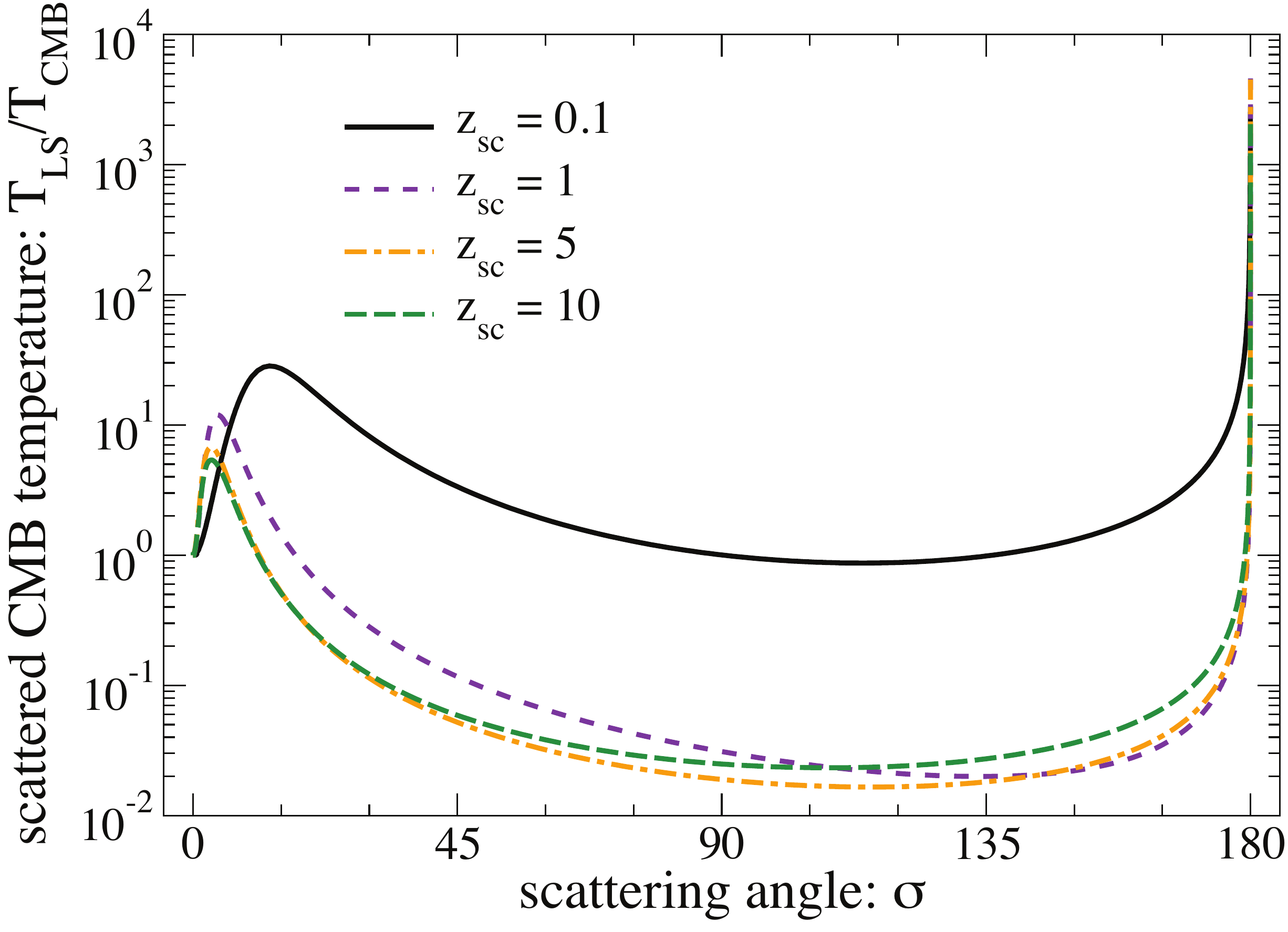}
\caption{The temperature $T_{LS}$ of secondary photons that scatter into the past light cone at a redshift $z_{sc}$ from a direction $\sigma$, relative to the temperature $T_{CMB}$ of primary photons traveling on the past light cone, are shown. Photons at $\sim 15^\circ,\, 180^\circ$ are substantially hotter than unscattered photons, despite the fact that all photons originate from the same constant density surface. This is a consequence of a negative radial expansion rate on a portion of this surface, which leads to a strong blueshifting of the photon energies.}
\label{fig:zcrshsigma}
\end{figure}

To develop some understanding of the behavior of geodesics inside the past light cone, we present a few results of the numerical integration of the geodesic equations. In Fig.~\ref{fig:180rescatter} we show a series of geodesics in which the scattered segment is $180^\circ$ opposite the past light cone. Although the past light cone lies along a straight line in the $t-r$ plane, the scattered geodesics are slightly curved. Geodesics that scatter at redshift $z \lesssim 1$ return back through the origin. The geodesics pictured run until decoupling. In a homogeneous universe we would define decoupling by the condition $\Gamma = H$, where $\Gamma$ is the photon-electron scattering interaction rate and $H$ is the expansion scale factor. In our case, because there is not a unique expansion rate, we define decoupling as the surface at which the matter density reaches a value $(1+z_{CMB})^3$ times the present-day density, where $z_{CMB} = 1100$. If we call the redshift of such a geodesic $z_{LS}$, then $z_{LS} \neq z_{CMB}$ in general. To illustrate this point, in Fig.~\ref{fig:zcrshsigma} we show the temperature of scattered photons traveling along the path ${BSO}$ relative to the temperature of unscattered, primary photons. The curves show the dependence on scattering angle and scattering redshift. For the range of redshifts shown, scattering at $\sim 15^\circ,\, 180^\circ$ produces surprisingly large photon temperatures or energies that will contribute to a large u-distortion. As we explain later, in Sec.~\ref{sec:pathology}, the large temperatures arise because of the strong gravitational fields along the geodesic paths inside the past light cone.

\section{U Distortion}
\label{sec:udist}

Our toy model does not contain enough physics to properly describe the CMB. The space-time contains matter and no radiation, and there is no theory of initial conditions for the fluctuations that would explain the observed temperature anisotropy pattern \cite{Bennett:2012zja,Planck:2013kta}. However, since decoupling occurs within the matter-dominated era in the Standard Cosmological Model, we feel justified to ignore the effect of radiation on the expansion history. Next, the radial inhomogeneity of the space-time is a sufficient source of spectral distortion, as will be shown, without needing a theory of primordial fluctuations. Hence, we model CMB photons by tracing geodesics backwards from the present-day origin until the energy density reaches a value $(1+z_{CMB})^3$ times the present-day value, where $z_{CMB} = 1100$. We assume that the CMB photons at this surface of constant matter density is a pure blackbody of temperature $T_{CMB} = T_0 \times (1+z_{CMB})$. By the present day, this CMB is isotropic on the sky, but no longer a blackbody.
 
The u-distortion is the leading contribution to the spectral distortion of the CMB blackbody due to the mixing of two or more blackbodies at slightly different temperatures (e.g. Refs.~\cite{Zeldovich:1972,Chluba:2004cn,Stebbins:2007ve,Khatri:2012rt}. As an example, consider two blackbodies with temperatures $T$ and $T + \Delta T$, mixed with weights $1-w$ and $w$ respectively, where $w \ll 1$. The resulting shift in the spectral intensity, relative to the intensity of a reference blackbody at temperature $\bar T = T + w \Delta T$, is given by
\begin{eqnarray}
\Delta I &=& \big( (1-w)I(T) + w I(T+\Delta T)\big) - I(\bar T) \cr 
&\simeq & \frac{1}{2} u \, T^2 I''(T)
\end{eqnarray}
where $u = w (\Delta T/T)^2$ and the prime indicates a derivative with respect to temperature. Because $w \Delta T \ll 1$, we are justified to make this expansion. Since the Compton y-distortion of the CMB takes the same form, $\Delta I = y \, T^2 I''$, we can equate $u = 2 y$ to translate the observational limit on $y$ into a limit on $u$. For the continuous mixture of blackbodies, due to CMB photons that scatter into our line of sight, the weight becomes an integral along the past light cone, over all single scattering directions that mix blackbodies, whereby the u-distortion is
\begin{eqnarray}
u &=& \int d\lambda \, n_e \int d\hat n'  \left( \frac{d\sigma_T}{d\Omega} \right) \left[ \frac{\Delta T}{T}(\lambda,\hat n') - \frac{\Delta T}{T}(\lambda,\hat n)\right]^2  \cr
&=& \frac{3 \sigma_T}{16 \pi}  \int d\lambda \, n_e \int d\hat n' \, \left(1 + (\hat n \cdot \hat n')^2 \right) \left[ \frac{\Delta T}{T}(\lambda,\hat n') - \frac{\Delta T}{T}(\lambda,\hat n)\right]^2.
\label{eqn:udist}
\end{eqnarray}
Here $d\sigma_T/d\Omega$ is the differential Thomson cross section, $n_e$ is the free electron density, and $\lambda$ is the length of the photon's path. We define $\mu = \hat n \cdot \hat n'$, the cosine of the angle between the scattered photon and the line of sight, so that $\mu=1$ corresponds to no change in direction, and $\mu=-1$ is scattering in the opposite direction. We ignore the small perturbations that are responsible for the standard CMB anisotropy, whereby ${\Delta T}(\lambda,\hat n) = 0$. However, the temperature in a direction $\hat n'$ is $T(\lambda,\hat n') = T_{CMB}/(1 + z_{LS}(\lambda,\hat n'))$ so that
\begin{equation}
\frac{\Delta T}{T}(\lambda,\hat n') = \frac{1+z_{CMB}}{1+ z_{LS}(\lambda,\hat n')}-1.
\end{equation}
A common assumption is that the temperature pattern is well approximated by a dipole, although a quick glance at Figure~\ref{fig:zcrshsigma} should convince the reader that this is not applicable in our scenario. To isolate the angular portion of the integral (\ref{eqn:udist}), we define
\begin{eqnarray}
I(z) &=& \int d\hat n' \, \left(1 + (\hat n \cdot \hat n')^2 \right) \left[ \frac{\Delta T}{T}(\lambda,\hat n') - \frac{\Delta T}{T}(\lambda,\hat n)\right]^2 \cr
&=& 2 \pi \int_{-1}^{1} d\mu \, (1 + \mu^2) \left[ \frac{1+z_{CMB}}{1 + z_{LS}(z,\mu)} - 1\right]^2.
\end{eqnarray}
Then the full expression for the u-distortion is
\begin{equation}
u  = \frac{3 \sigma_T}{16 \pi} \left( \frac{1 - \frac{1}{2}Y_{He}}{m_H}\right)   \left( 3 \Omega_B H_0^2 c/\kappa \right) \int_0^{z_R} dz\, \frac{(1+z)^2}{H_{\Lambda CDM}(z)} I(z).
\end{equation}
The range of integration is determined by the optical depth to the period of reionization,
\begin{equation}
\tau = \int d\lambda \,  n_e \sigma_T.
\end{equation}
We assume that the number density of electrons tracks the matter density along the past light cone, and that reionization takes place ``suddenly". In order that $\tau \simeq 0.09$, in agreement with WMAP \cite{Bennett:2012fp} and Planck \cite{Ade:2013zuv}, with parameters $\Omega_B h^2 = 0.022$ and $h=0.7$, we set the redshift of reionization as $z_R \simeq 10$. 

As the redshift of rescatter decreases, the photons grow increasingly energetic. Therefore the u-distortion is mildly sensitive to the lower limit of integration; likewise, at the lower limit our idealistic description of a smooth mass distribution breaks down. In order to account for the transition to homogeneity and the paucity of scattering electrons along any particular line of sight, due to the clustering of large scale structure, we weight the number density of scatterers as in Ref.~\cite{Scrimgeour:2012wt} by
\begin{equation}
n = N/V \to N / \int dV (1 + \xi(s)).
\end{equation}
In the above equation $dV = 4 \pi s^2 ds$, $\xi = (s_0/s)^\gamma$ is the clustering correlation function, with $\gamma = 1.8$, and we use a comoving length $s_0 = 20-60$~Mpc/h, characteristic of a cluster or supercluster length scale (e.g. Refs.~\cite{Bahcall:1986,Basilakos:2003ux,Bahcall:2003hu}). Hence, we insert a window function into the integrand for the u-distortion,
\begin{equation}
W(s[z]) = s^\gamma /( s^\gamma + 3 s_0^\gamma/(3-\gamma) )
\end{equation}
where $s[z] \simeq \widehat R$ for the radii of interest \cite{Scrimgeour:2012wt}. Essentially, this window function introduces a lower limit of integration $z_{min} \simeq 0.02 - 0.05$. To be conservative, we use the longest clustering scale, although the result varies by less than 10\%. 

Our final result for the u-distortion is $u = 0.11$. Compared to the COBE-FIRAS limit $u < 3.0 \times 10^{-5}$ (95\% CL) \cite{Mather:1993ij,Fixsen:1996nj}, our model predicts a signal that exceeds the observational bound by a factor of 3700. This simple model is clearly in conflict with observational evidence.

The width of the temperature distribution, as defined in Ref.~\cite{Stebbins:2007ve}, replaces the $\Delta T/T$-squared term in Eq.~\ref{eqn:udist} by $log(1 + \Delta T/T)$-squared. For small temperature differences, these are equivalent. As we have seen in this scenario, the temperature of scattered photons that originate from within the past light cone can be substantially different from those on the past light cone: $\Delta T/T \not\ll 1$. Using the log term to evaluate the u-distortion, we obtain an even larger result: $u = 0.90$. This may seem surprising, since taking the log might be expected to suppress the effect of large values of $\Delta T/T$, but that is only true when $\Delta T/T$ is positive. For sufficiently negative values of $\Delta T/T$, when cold photons are injected into the observed stream and CMB photons are scattered out, the argument of the log term approaches zero and the integrand becomes large. In our calculation, we use the $\Delta T/T$-squared term because it is closer to the procedure carried out by the COBE-FIRAS team.

\section{Strange Properties of the Space-Time}
\label{sec:pathology}

The large amplitude of the u-distortion strongly suggests that the space-time suffers from a number of pathologies. In this section we investigate some of these features, in part out of curiosity and in part with foresight to the analysis of similar models that attempt to dispense with dark energy.

\subsection{Radial Contraction}

A surprising feature of this toy model is the presence of a radially contracting region inside the past light cone. Photons originating from the epoch of decoupling pass through this region and gain energy, so that consequently upon scattering into our line of sight, they contribute to the strong u-distortion. The redshifting of radially-directed photons is guided by $H_{\parallel}$. Although the radial expansion rate matches the $\Lambda$CDM expansion rate on the past light cone, inside the past light cone the behavior of $H_\parallel$ is dramatically different.

\begin{figure}[h]
\includegraphics[width=0.5\linewidth]{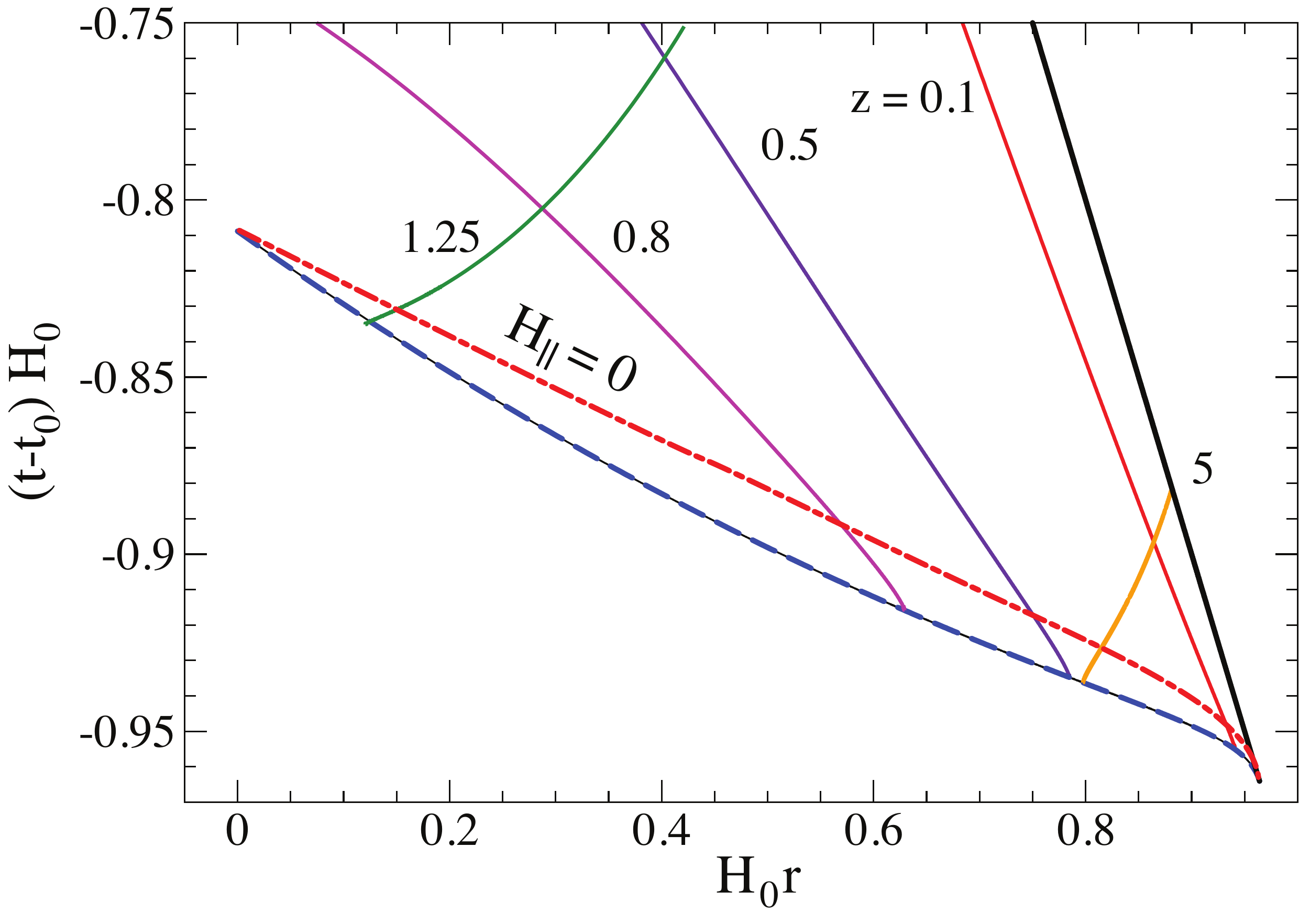}
\caption{A space-time diagram showing the surface where the radial contraction turns to expansion ($H_\parallel =0$; thick red dot-dashed line) and the constant energy density surface at decoupling (thin black line). The same scattered geodesics as in Fig.~\ref{fig:180rescatter} are also shown. All lines are shown with $r\ge 0$. The decoupling contour and the Big Bang surface (thick dashed blue lines) cannot be distinguished by eye in this figure.}
\label{fig:zturn}
\end{figure}

The matter density along the $H_\parallel = 0$ surface, which is illustrated in Fig.~\ref{fig:zturn}, is less than the critical density of decoupling for a wide range of radii. This means that a large portion of the photons that scatter into our line of sight must pass through the $H_\parallel < 0$ region. By inspecting Eq.~(\ref{Ceqn:dzdt}) for the growth of redshift $z$ along a photon's trajectory, we infer that when $H_\parallel < 0$, radially-directed photons gain energy. The general condition for the photon to gain energy is $H_\parallel < - H_\perp \tan^2\psi$, where $\sin\psi = \ell/(1+z)R$. Hence, some of the light rays that are emitted from the decoupling epoch and subsequently scatter into our past light cone are much more energetic, with a much lower redshift $z_{LS} \ll z_{CMB}$, than photons that have traveled along the past light cone since decoupling. This feature of the inhomogeneous Big Bang LTB space-time, also discussed in Refs.~\cite{Bull:2011wi,Zibin:2011ma}, is ultimately responsible for the large u-distortion signal that we calculate.

\subsection{Naked Singularity} 

The contracting region continues all the way down to the Big Bang surface, leading to the perverse situation that the Big Bang singularity is accessible to photons at a finite redshift. In a Robertson-Walker space-time, light rays emanating from the Big Bang that propagate along our past light cone are infinitely redshifted with $z \to \infty$. In our toy model, light rays that start at the Big Bang and propagate along the past light cone of here and now are also infinitely redshifted. However, light rays that begin at the Big Bang from within our past light cone and scatter onto our line of sight will be observed with redshift $z=-1$, as explored in Ref.~\cite{Hellaby:1984}.

The Big Bang is defined by the surface $\eta(r,\,t)=0$ in Eqs.~(\ref{eqn:param1}-\ref{eqn:param2}). Approaching this surface, as $\eta \to 0$, then $R \simeq {\alpha  \eta^2 }/{4 \beta}$. Since $\alpha$ and $\beta$ are finite at a given value of $r$, then we can use Eq.~(\ref{eqn:EEeqn1}) to determine that $H_\perp = \dot R/R \to \infty$ as $\eta\to 0$ at the Big Bang. The expansion rate $H_\parallel$ is different. To understand its behavior, we need to know the $\eta$-dependence of $R'$ near the Big Bang. We can do so by following the calculations of KL \cite{Kolb:2009hn}. Along fixed $r$, dropping down in $t$ towards the Big Bang, as shown in Fig.~\ref{fig:symbolic}, then Eq.~(\ref{eqn:EEeqn2}) tells us that the rate of change in $R'$ is
\begin{eqnarray}
\frac{\partial}{\partial \eta}R' &=& \dot R'   \frac{\partial t}{\partial \eta} \cr
&=& \frac{\beta' + \alpha'/R - \alpha R'/R^2}{2 \dot R} \frac{R}{\sqrt{\beta}} \cr
& \simeq & \frac{\alpha'}{4 \beta} \eta - R' / \eta .
\end{eqnarray}
The last line is obtained by taking the approximate behavior near $\eta \to 0$. The solution of this differential equation is 
\begin{equation}
R' \simeq \frac{\alpha'}{12 \beta}\eta^2 - 2 \sqrt{\beta} t_{BB}' /\eta .
\end{equation}
Since the Big Bang surface is inhomogeneous, $\partial_r t_{BB} \neq 0$, then $R'$ is singular. Put together, then in the vicinity of the Big Bang surface,
\begin{equation}
H_\parallel = \dot R' /R' = \frac{d\eta}{dt} \frac{\partial R'}{\partial \eta} / R'  \simeq - \frac{\beta}{\eta R} \propto \eta^{-3}.
\end{equation}
So the radial expansion rate is negative and divergent as $\eta \to 0$.

The geodesic equation for the evolution of redshift, $z$, inside the light cone is given by Eq.~(\ref{Ceqn:dzdt}). If we consider a radially-directed light ray that starts at an earlier and earlier instant close to the Big Bang, then $\ln(1+z)$ receives a successively negative and divergent contribution, such that $z\to -1$. This constitutes a naked singularity: there is no horizon to protect us from the singular curvature of the Big Bang.

\subsection{Anisotropic Expansion}

\begin{figure}[h]
\includegraphics[width=0.45\linewidth]{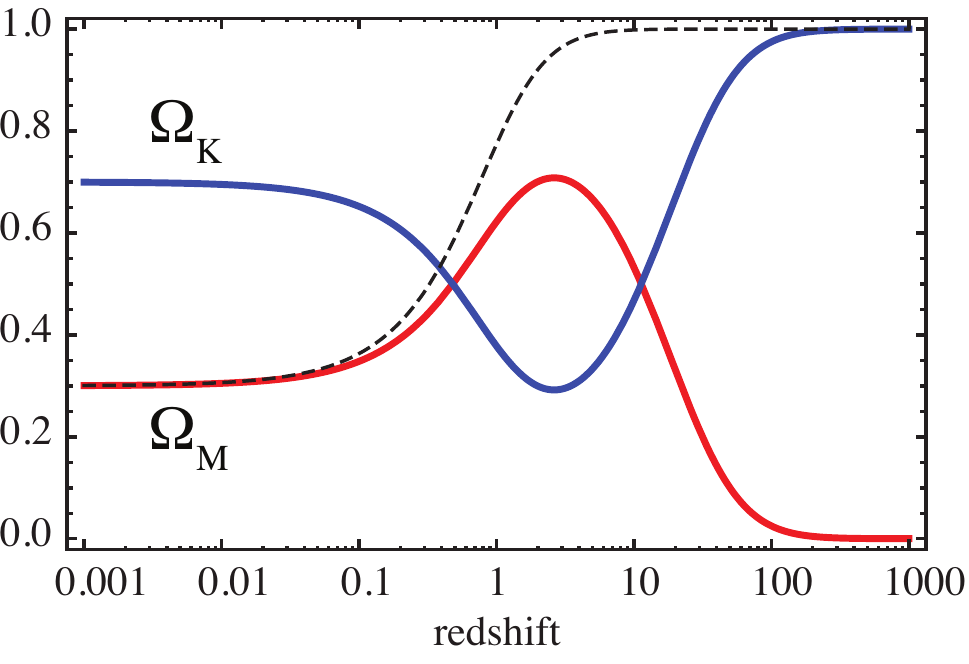}
\hspace{0.2cm}
\includegraphics[width=0.49\linewidth]{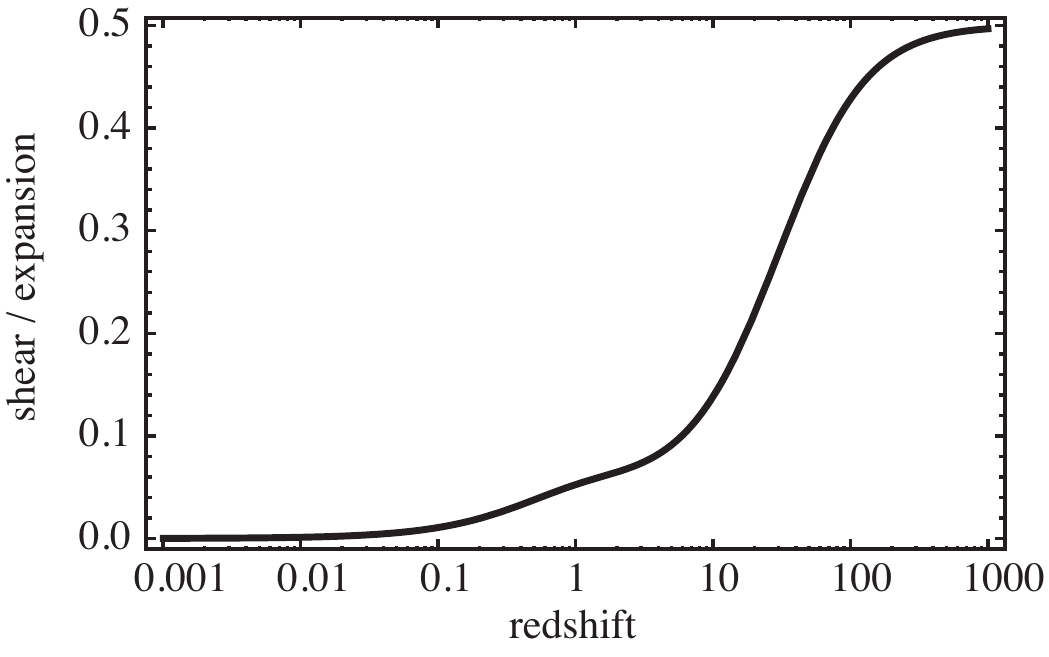}
\caption{(left) The spatial curvature and density in units of the anisotropic expansion rate along the past light cone. The dashed line shows the density in units of the radial expansion rate. (right) The ratio of the shear to the expansion, $(H_\perp- H_\parallel)/(2 H_\perp + H_\parallel)$, as a measure of the anisotropy of the expansion.}
\label{fig:kcurv}
\end{figure}

Another of the many peculiar features of this space-time is the anisotropic expansion along the past light cone of here and now. To see this, let us define
\begin{eqnarray}
\Omega_M(z) &=& \kappa\rho / (2 H_\perp H_\parallel + H_\perp^2) \cr
\Omega_K(z) &=& {}^{(3)} R/ (2 H_\perp H_\parallel + H_\perp^2) \cr
{}^{(3)}R &=& -{(\beta R)'}/{R^2 R'}
\end{eqnarray}
in which case the Einstein field equation (\ref{eqn:gtt}) can be written as a familiar sum law, $\Omega_M + \Omega_K = 1$. The behavior of these two parameters is different from that of the standard, Robertson-Walker space-time. Although they take the expected values $0.3,\, 0.7$ at the present day, the values at deep redshift are different: $\Omega_M \to 0,\, \Omega_K \to 1$. The expansion grows increasingly anisotropic with redshift. Although the expansion along the radial direction matches that of a $\Lambda$CDM cosmology, with $H_\parallel = H_{\Lambda CDM}$ along the line of sight, the expansion in the transverse direction is faster, $H_\perp \ge H_\parallel$. To illustrate, in Fig.~\ref{fig:kcurv} we plot the redshift evolution of $\Omega_M,\, \Omega_K$ and $(H_\perp- H_\parallel)/(2 H_\perp + H_\parallel)$, which is proportional to the ratio of the shear to the expansion. We note that the transverse length scales of the baryon acoustic oscillations (BAO) observed in galaxy clustering patterns \cite{Eisenstein:2005su,Percival:2007yw,Percival:2009xn,Blake:2011en} are sensitive to the deviation from isotropic expansion, i.e. the departure from unity by $H_\perp / H_\parallel$. In our model, however, this is less than a $20\%$ effect for $z<1$, so that the comparison with BAO data may not be too discrepant. This puts tension on the viability of our toy model similar to the effects investigated in Ref.~\cite{Zumalacarregui:2012pq}, though nowhere near as strong as that due to the u-distortion.

\subsection{Locating the Big Bang surface}

An inhomogeneous Big Bang surface implies that the age of the Universe at a given time varies with radius. 
The time since the Big Bang can be calculated at any point using Eq.~(\ref{eqn:param2}). 
In the vicinity of the origin, for $r H_0 \ll 1$, we can use $\alpha \simeq \Omega H_0^2 r^3$, $\beta \simeq (1-\Omega) H_0^2 r^2$ and $\widehat R \simeq r$ to determine $\eta(t_0,0) = \cosh^{-1}\left( 1 - {2}/{\Omega}\right).$ In turn, this yields an age of the Universe at the origin 
\begin{equation}
t_0 - t_{BB}(0) = H_0^{-1}\left( \frac{1}{1- \Omega} - \frac{\Omega}{2(1-\Omega)^{3/2}}\cosh^{-1}\left( 1 - \frac{2}{\Omega}\right) \right).
\end{equation}
Using our standard values $\Omega = 0.3$ and $H_0 = 70$~km/s/Mpc, this yields $11.3$~Gyrs. This is close to the $11.2$~Gyr, $95\%$~CL lower limit on the age of the Universe based on Milky Way globular clusters, although an additional $0.1 - 2$~Gyrs must also be allotted for the formation time of the stars in the galactic halo \cite{Krauss:2003em}. In order to accommodate these old stars in our toy model, we may either lower the Hubble constant $H_0$ or the matter density $\Omega$ in order to achieve an adequate time since the Big Bang.  

The shape of the inhomogeneous Big Bang surface is shown in earlier figures. The principle features of the curve include a cusp at the origin and a downturn at $r \sim 0.9$. 
We note that these features are distinct from the Gaussian profile for $t_{BB}(r)$ that is assumed in Ref.~\cite{Zibin:2011ma}.  

At the edge of the observable Universe, $r_{max} = \frac{2}{3} H_0^{-1}\tanh^{-1}\sqrt{1-\Omega} / \sqrt{1-\Omega}$. Evaluating the time of the Big Bang at this radius, we find that $t_{BB}$ is $\sim 2$~Gyrs earlier than at the origin. Of particular interest for the physics of the CMB in this model, the light travel time from the Big Bang to $z=1100$ is $\sim 0.5$~Myrs, whereas the time elapsed for an observer at rest since the Big Bang  is only $\sim 2500$~years.

\section{Summary}
\label{sec:summary}

Solutions to the physics of cosmic acceleration that dispense with dark energy and new gravitational physics are immensely appealing. In a number of recent papers, it has been shown that the expansion history along the past light cone---determined by the luminosity distance, energy density, and number counts---can be built from the LTB metric to match the $\Lambda$CDM Standard Cosmological Model without the presence of a Gigaparsec void \cite{Celerier:2009sv,Kolb:2009hn,Dunsby:2010ts}. That is, a post-decoupling cosmos containing just dark matter and baryons obeying the laws of general relativity can satisfy many of the classical tests of cosmology. The price to pay seems philosophical, since for this to work we must be located at the center of a radially inhomogeneous space-time, contrary to the Copernican and Cosmological Principles. However, these LTB-based models must also pass a battery of other observational tests before we can discuss the possible meaning.

Our interest is to probe the space-time inside the past light cone, to dig beneath the surface. The u-distortion is among a handful of cosmological probes, along with the kinetic Sunyaev-Zeldovich effect \cite{GarciaBellido:2008gd,Yoo:2010ad,Zhang:2010fa,Bull:2011wi} and redshift drift \cite{Uzan:2008qp,Clifton:2008hv,Quartin:2009xr,Yoo:2010hi}, that have been pursued recently for their ability to probe below the past light cone and test radial inhomogeneity. Given a choice, experience shows that any constraint having to do with the CMB is usually the strongest. The u-distortion has previously proven useful in constraining constant-time Big Bang models \cite{Caldwell:2007yu,Moss:2010jx,Caldwell:2010}.  

In this paper we have carried out the first calculation of the u-distortion in a LTB model with a radially inhomogeneous Big Bang.  One paper in particular that influenced our work is KL \cite{Kolb:2009hn}, which clearly spelled out how to build such a space-time. Its influence on our work is clear---we adopt the same notation, reproduce several figures, and extend some of the calculations in KL. Our focus, therefore, has been to evaluate the u-distortion for the model described in KL. 

To recap this effect, the u-distortion quantifies the departure from a blackbody at temperature $T$ when additional blackbodies at temperature $T+\Delta T$ are added. This situation applies to the blackbody comprised of unscattered CMB photons that travel direct on our line of sight from decoupling, mixed with CMB photons that Thomson scatter into our line of sight. Since the energies of these two sets of photons are different in general, we can expect the observed blackbody to display a wide temperature distribution, or a significant u-distortion. Measurements by the COBE-FIRAS experiment place a tight limit, $u < 3.0 \times 10^{-5}$ (95\% CL) \cite{Mather:1993ij,Fixsen:1996nj}, where $u=2y$ at leading order, and $y$ is the Compton-y parameter. We find that the predicted value $u = 0.11$, exceeds the upper bound by a factor of $3700$. Without a doubt, the u-distortion is a decisive probe of radial inhomogeneity, as this model is ruled out.

A similar calculation was carried out in Ref.~\cite{Zibin:2011ma}, in which the u-distortion was evaluated for an LTB space-time with a Big Bang surface described by a Gaussian profile of variable amplitude and radius. We note that this work considers void models with whereas our scenario has no such void. Also, the Gaussian profile does not give a good fit to the Big Bang surface in our scenario. In particular, it does not ensure a past light cone history that so precisely matches the $\Lambda$CDM model. Nevertheless, Ref.~\cite{Zibin:2011ma} likewise finds that an impermissibly large u-distortion is predicted for a range of parameter values. (See Figs.~2, 3 therein.) Ref.~\cite{Zibin:2011ma} also highlights the shortcoming of the dipole approximation for the temperature anisotropy pattern. Overall, our results confirm the argument put forward in Ref.~\cite{Zibin:2011ma} that radial inhomogeneity of the Big Bang surface, alternatively referred to as decaying modes, cannot salvage a scenario based upon the spherically symmetric, dust-filled LTB models.

Is there any future for such models? It is conceivable that a more realistic treatment of decoupling and the radiation-dominated epoch could weaken the level of spectral distortion that we calculate. We have made the simplifying assumption of tight coupling of the radiation with baryonic and dark matter until decoupling, and adapted the Gamow criterion in order to identify the origin of the CMB with a critical value of the matter density (e.g. Refs.~\cite{Yoo:2010ad,Zibin:2011ma}). We implicitly assume that any slip that develops between the matter species and radiation leads to a negligible source of temperature anisotropy in our calculation of the spectral distortion. That this assumption may not be wholly justified has been argued in Ref.~\cite{Clarkson:2010ej}. We leave a more sophisticated treatment of the CMB for future work.
 
The surprising features of the space-time that we explore in Sec.~\ref{sec:pathology} are known in the literature. For example, the radial contraction was discussed in Refs.~\cite{Bull:2011wi,Zibin:2011ma}, wherein cosmological constraints were applied to a set of LTB models with a radially inhomogeneous Big Bang. The LTB models are well known to admit naked singularities, from investigations of the gravitational collapse of dust shells \cite{Eardley:1978tr,Christodoulou:1984mz}. The existence of infinitely blueshifted geodesics in the cosmological LTB metric was investigated in Ref.~\cite{Hellaby:1984}. If any aspect of this model is to survive as a viable alternative to the Standard Cosmological Model, then a large portion of the space-time inside the past light cone, including regions after decoupling, would have to be excised and replaced with a safer, less inhomogeneous space-time. However, the full implication of these phenomena have not been widely exploited in the development of cosmological probes. Possibly, the fact that everything appears standard on the past light cone of here and now makes the space-time seem safe to the innocent bystander. In view of the tremendous interest in testing alternatives to dark energy, radial homogeneity, and the Copernican Principle \cite{Clifton:2011sn,Clarkson:2012bg,Valkenburg:2012td,Bull:2013fga}, or just to determine how much information can actually be gained about our space-time \cite{Stebbins:2012vw,Stebbins:2012}, one of our motivations has been to demonstrate the efficacy of  spectral distortions to dig beneath the surface of a cosmological model.

\appendix

\section{Geodesics on the Past Light Cone}
\label{appendixA}

We consider the four-momentum of a photon $k^\mu = dx^\mu/d\lambda$ where $\lambda$ is an affine parameter that increases with time. In the case of a radially-directed light ray, the null condition $k \cdot k=0$ gives
\begin{equation}
\frac{dt}{d\lambda} = -\frac{R'(r,t)}{\sqrt{1 + \beta(r)}} \frac{dr}{d\lambda}
\label{Aeqn:dtdl}
\end{equation}
where the sign is chosen for a light ray moving towards decreasing $r$, for $R' > 0$ (which is the case). The energy of the light ray, as determined by a comoving observer with four-velocity $u^\mu = (1,0,0,0)$ is ${\cal E} = - u \cdot k$. Consequently, we define the redshift to be the ratio of the initial energy at point $A$ to the final energy at point $O$ as illustrated in Fig.~\ref{fig:spacetime}
\begin{equation}
1+z = \frac{u \cdot k|_A}{u \cdot k|_O}.
\end{equation}
If we rescale $\lambda$ so that $dt/d\lambda |_O=1$, then we have the equation for redshift
\begin{equation}
\frac{dt}{d\lambda} = 1+z.
\end{equation}
Next we can use the geodesic equation to determine the evolution of $z$:
\begin{eqnarray}
&& \frac{d k^\mu}{d\lambda} + \Gamma^\mu_{\alpha\beta} k^\alpha k^\beta = 0 \cr
&& {\mu=t}: \quad \frac{dz}{d\lambda} + \frac{\dot R'(r,t)}{R'(r,t)}(1+z)^2 =0.
\label{Aeqn:geodesic}
\end{eqnarray}
We can use Eqs.~(\ref{Aeqn:dtdl}, \ref{Aeqn:geodesic}) to arrive at the desired results,
\begin{eqnarray}
\frac{dr}{dt} &=& - \frac{\sqrt{1 + \beta(r)}}{R'(r,t)} \\
\frac{dz}{dt} &=& - \frac{\dot R'(r,t)}{R'(r,t)}(1+z).
\end{eqnarray}
Using the coordinate freedom to rescale the radial coordinate $r$, whereby $R' = \sqrt{1+\beta}$ on the past light cone, then $dr/dt = -1$.

\section{Calculations on the Past Light Cone}
\label{appendixB}

Our procedure for evaluating $\alpha,\, \beta$ and radial derivatives follows that of KL. To derive a differential equation for $\alpha$, we start from the definition of energy density whereby $\alpha'(r) = \kappa \rho R^2 R' $. On the past light cone $\widehat R' = \sqrt{1 + \beta}$. Squaring both sides, and using
\begin{equation}
\frac{d \widehat R}{dr} = \widehat R' - \partial_t{\widehat R},
\end{equation}
we obtain
\begin{equation}
\widehat R' \frac{d\widehat R}{dr} = 1 + \beta - \widehat R' \partial_t{\widehat R}.
\end{equation}
Next, using Eq.~(\ref{eqn:EEeqn1}) to replace $\beta$, we can write
\begin{eqnarray}
\widehat R' \frac{d\widehat R}{dr} &=& 1 - \frac{\alpha}{\widehat R} + \partial_t{\widehat R}^2 -  \widehat R' \partial_t{\widehat R}, \cr
                                                      &=& 1 - \frac{\alpha}{\widehat R} + \partial_t{\widehat R} \left(  \partial_t{\widehat R}-  \widehat R' \right), \cr
                                                      &=& 1 - \frac{\alpha}{\widehat R} + \left( \frac{d\widehat R}{dr} - \widehat R'\right) \left( \frac{d\widehat R}{dr} \right), \cr
2\widehat R' \frac{d\widehat R}{dr} &=& 1 - \frac{\alpha}{\widehat R}+\left( \frac{d\widehat R}{dr} \right)^2 \cr      
 \widehat R'  &=& \frac{1}{2}\left[ \left(1 - \frac{\alpha}{\widehat R}\right)/\left( \frac{d\widehat R}{dr}\right)+\frac{d\widehat R}{dr} \right] .
\end{eqnarray}
Now we return to the energy density equation, to obtain
\begin{equation}
\alpha'(r) = \frac{1}{2}\kappa \widehat\rho \widehat R^2 
\left[ \left(1 - \frac{\alpha}{\widehat R}\right)/\left( \frac{d\widehat R}{dr}\right)+\frac{d\widehat R}{dr} \right]. 
\label{Beqn:dadr}
\end{equation}
In practice, the integration along the past light cone is carried out with respect to $z$, not $r$, so instead we use
\begin{equation}
\frac{d\alpha}{dz} =  \frac{1}{2}\kappa \widehat\rho \widehat R^2 
\left[ \left(1 - \frac{\alpha}{\widehat R}\right)\left(\frac{dr}{dz}\right)^2/\left( \frac{d\widehat R}{dz}\right)+\frac{d\widehat R}{dz} \right]. 
\label{Beqn:dadz}
\end{equation}
where $dz/dr = (1+z)H_{\Lambda CDM}(z)$ and $d\widehat R/dz = -(\widehat R - 1/H_{\Lambda CDM}(z))/(1+z)$.
The equation for $\beta$ is obtained by squaring $\widehat R' = \sqrt{1 + \beta}$ and using the expression for energy density:
\begin{eqnarray}
\sqrt{1 + \beta} &=& \widehat R' \cr
1+\beta &=& \widehat R'^2 \cr
\beta(r) &=&  \left(\frac{\alpha'(r)}{\kappa \widehat \rho \widehat R^2}\right)^2 -1.
\label{Beqn:beta}
\end{eqnarray}
By substituting $\alpha'(r)$ from Eq.~(\ref{Beqn:dadr}) into the above, we can then differentiate $\beta$ to obtain $\beta'(r)$. An equivalent expression can be obtained by noting that Eqs.~(\ref{eqn:drdt}-\ref{eqn:dzdt}) indicate that $\dot R'/R'=H_{\Lambda CDM}(z)$ on the past light cone. Using the Einstein field equation (\ref{eqn:EEeqn2}) for $\dot R'$, we arrive at
\begin{equation}
\beta' = -\alpha'/\widehat R + \alpha \sqrt{1+\beta} / \widehat R^2 + 2 H_{\Lambda CDM}\sqrt{\left(1+\beta\right)\left(\beta + \alpha/\widehat R\right)}. 
\label{Beqn:dbdr}
\end{equation}
Eqs.~(\ref{Beqn:dadz}-\ref{Beqn:dbdr}) along with Eq.~(\ref{eqn:dzdr}) provide all the ingredients needed to construct $\alpha$, $\beta$ and derivatives for any value of the radial coordinate.

It is difficult to numerically evaluate $\alpha,\,\beta$ in the vicinity of $z \sim 1.6$ for our chosen value $\Omega = 0.3$. It is near this redshift that $\widehat R = H_{\Lambda CDM}^{-1}$, which means
\begin{equation}
\frac{d\widehat R}{dr} = 1 - \widehat R H_{\Lambda CDM}
\end{equation}
vanishes. Fortunately, $\alpha = \widehat R$ at the same location, so that the equation for $\alpha'$ is not singular. However, numerical evolution of the differential equation for $\alpha$ becomes challenging. Our strategy is to define a variable 
\begin{equation}
V \equiv \frac{1 - \alpha / \widehat R}{1 - \widehat R H_{\Lambda CDM}}
\end{equation}
and to rewrite the equations for $\alpha,\,\beta$ using the explicit form of $H_{\Lambda CDM}$.
\begin{eqnarray}
\frac{dV}{dz} &=& \frac{1 - \frac{3}{2}\Omega (1+z)^3 \widehat R^2 H_0^2 + V (2 \widehat R H_{\Lambda CDM} - 1)}{(1+z) \widehat R H_{\Lambda CDM} }
\label{Beqn:dVdz} \\
\alpha &=& \widehat R (1 - V (1 - \widehat R H_{\Lambda CDM})) \\
\beta &=& \frac{1}{4}(1 - \widehat R H_{\Lambda CDM} + V)^2 - 1 \\
\alpha' &=& \frac{3}{2} \Omega H_0^2(1+z)^3 \widehat R^2 (1 +V - \widehat R H_{\Lambda CDM}) \\
\beta' &=& \frac{1}{2 \widehat R}(1 +V - \widehat R H_{\Lambda CDM})\left[
1 - V+ 2 V \widehat R H_{\Lambda CDM} - \widehat R H_{\Lambda CDM}+ (\widehat R H_{\Lambda CDM})^2 - 3 \Omega H_0^2 (1+z)^3 \widehat R^2 \right] .
\label{Beqn:lasteqn}
\end{eqnarray}
This formulation of the equations is well behaved across $z \sim 1.6$, so we need only to integrate (\ref{Beqn:dVdz}) and use Eqs.~(\ref{eqn:Rofz}, \ref{eqn:dzdr}) to construct  $\alpha,\,\beta$ as functions of $r$. Near the origin, for $r H_0\ll 1$,  our recipe yields $\alpha \simeq \Omega H_0^2 r^3 + {\cal O}(r^4)$, $\beta \simeq (1-\Omega) H_0^2 r^2  + {\cal O}(r^3)$, $V \simeq 1 + r H_0 + {\cal O}(r^2)$, $\widehat R \simeq r  + {\cal O}(r^2)$, and $r \simeq H_0^{-1} z + {\cal O}(z^2)$.

\section{Geodesics off the Past Light Cone}
\label{appendixC}

We consider the four-momentum of a photon $k^\mu = dx^\mu/d\lambda$ where $\lambda$ is an affine parameter that increases with time. In the case of a light ray moving in the equatorial plane ($\theta = \pi/2$), the null condition $k \cdot k=0$ gives
\begin{equation}
\left(\frac{dt}{d\lambda}\right)^2 = \frac{R'^2(r,t)}{1 + \beta(r)} \left(\frac{dr}{d\lambda}\right)^2 + R^2(r,t) \left( \frac{d\phi}{d\lambda} \right)^2.
\label{Ceqn:dtdl}
\end{equation}
The space-time has a Killing vector oriented in the $\phi$-direction, so that $\phi$-motion is conserved whereby
\begin{equation}
R^2(r,t)\frac{d\phi}{d\lambda} = \ell
\label{Ceqn:ell}
\end{equation}
such that $\ell$ is a constant of motion, the angular momentum per unit energy. As before, the energy of the light ray, as determined by a comoving observer with four-velocity $u^\mu = (1,0,0,0)$ is ${\cal E} = - u \cdot k$. Consequently, we again obtain the equation for redshift
\begin{equation}
\frac{dt}{d\lambda} = 1+z.
\end{equation}
The geodesic equation for the evolution of $z$ gives
\begin{equation}
\frac{d}{dt}\ln(1+z) = -\frac{\dot R'(r,t)}{R'(r,t)} + \left( \frac{\dot R'(r,t)}{R'(r,t)} - \frac{\dot R(r,t)}{R(r,t)} \right) \left[ \frac{\ell}{(1+z) R(r,t)}\right]^2.
\label{Ceqn:dzdt}
\end{equation}
Next, making the definition 
\begin{equation}
u(r,t) = (1+z) \frac{R'(r,t)}{\sqrt{1 + \beta(r)}} \frac{dr}{dt}
\label{Ceqn:udefn}
\end{equation}
the geodesic equation for $r$ can be expressed as
\begin{equation}
\frac{d u}{dt} = -\frac{\dot R'(r,t)}{R'(r,t)} u + \frac{ \sqrt{1+\beta(r)}}{(1+z) R(r,t)} \left( \frac{\ell}{R(r,t)} \right)^2.
\label{Ceqn:dudt}
\end{equation}
Eq.~(\ref{Ceqn:dtdl}) can also be manipulated to give a condition on $u$,
\begin{equation}
(1+z)^2 =u^2(r,t) + \left(\frac{\ell}{R(r,t)}\right)^2.
\label{Ceqn:ucon}
\end{equation}
The system of equations (\ref{Ceqn:dzdt}-\ref{Ceqn:dudt}) are sufficient to solve for general geodesic motion. 

To model a general geodesic originating at $B$ that scatters off an electron at $S$ on the past light cone of a present-day observer at $O$, it is useful to think about this process in reverse. We model a geodesic on the past light cone $OS$ to a location $(t_S,\,r_S)$ at which point $R=R_S$ and the redshift is $z_S$. Then, the geodesic continues in a new direction at an angle $\sigma \in [0,\pi)$. At the scattering site, the path from the origin points in a direction $\hat n_{OS}$ and the redirected path is $\hat n_{SB}$. In the local, comoving reference frame we define the angle $\sigma$ between the two paths by $\hat n_{OS} \cdot \hat n_{SB} = \cos\sigma$. The geodesic $SB$ has angular momentum parameter $\ell = R_{S} (1+z_S) \sin\sigma$ at which point $u = -(1+z_S)\cos\sigma$.

\acknowledgments
The work of RRC is supported in part by NSF PHY-1068027. The work of NAM is supported in part by the Hellman Family Foundation at Dartmouth College.
We thank Philip Bull, Jens Chluba, Chris Clarkson, Marco Regis, and James Zibin for useful comments.


\end{document}